\documentclass[aps,pre,showpacs,showkeys,preprint,floatfix,superscriptaddress]{revtex4-1}

\usepackage{amssymb}
\usepackage{amsmath}
\usepackage{amsbsy}
\usepackage{bm}
\usepackage{color}
\usepackage{graphicx}
\newcommand{\bq}{\begin{eqnarray}}
\newcommand{\eq}{\end{eqnarray}}
\newcommand{\bqn}{\begin{eqnarray*}}
\newcommand{\eqn}{\end{eqnarray*}}
\newcommand{\RR}{{\bf R}}
\newcommand{\rr}{{\bf r}}
\newcommand{\pp}{{\bf p}}
\newcommand{\PP}{{\bf P}}
\newcommand{\bb}{{\bf b}}
\newcommand{\BB}{{\bf B}}
\newcommand{\ff}{{\bf f}}
\newcommand{\FF}{{\bf F}}
\newcommand{\DD}{{\bf D}}
\newcommand{\nn}{{\bf n}}
\newcommand{\VV}{{\bf V}}
\newcommand{\ch}{{\cal H}}
\newcommand{\cl}{{\cal L}}
\newcommand{\cd}{{\cal D}}
\newcommand{\cs}{{\cal S}}
\newcommand{\ci}{\pmb{\cal I}}
\newcommand{\bnabla}{\pmb{\nabla}}
\newcommand{\pint}{\int^{\!\raisebox{-1.5pt}{\mbox{\textendash}}}_{\!\!\!\!\raisebox{-1pt}{\mbox{\textendash}}}\!\!\!\!\!\!\!\int}

\begin{document}
%%%%%%%%%%%%%%%%%%%%%%%%%%%%%%%%%%%%%%%%%%%%%%%%%%%%%%%%%%%%%%%%%%%%%%%%%%%%%%
%%%%%%%%%%%%%%%%%%%%%%%%%%%%%%%%%%%%%%%%%%%%%%%%%%%%%%%%%%%%%%%%%%%%%%%%%%%%%%
%%%%%%%%%%%%%%%%%%%%%%%%%%%%%%%%%%%%%%%%%%%%%%%%%%%%%%%%%%%%%%%%%%%%%%%%%%%%%%
\title{From the Liouville to the Smoluchowski equation for a colloidal
  solute particle in a solvent} 

\author{Riccardo Fantoni}
\email{rfantoni@ts.infn.it}
\affiliation{Universit\`a di Trieste, Dipartimento di Fisica, strada
  Costiera 11, 34151 Grignano (Trieste), Italy}

\date{\today}

\begin{abstract}
We show how the Smoluchowski dynamics of a colloidal Brownian particle
suspended in a molecular solvent can be reached starting from the
microscopic Liouvillian evolution of the full classical model in the
high friction limit. The integration of the solvent degrees of freedom
goes through a multiple time scale perturbation expansion which
removes the secular divergences. A simple dynamical Monte Carlo
scheme is then proposed to solve the resulting evolution equation for
the colloid solute particle. In particular we study the approach to
the equilibrium 
Boltzmann distribution at late times and its resilience behavior at
shorter times as influenced by the steepness of the external potential
and the friction coefficient around their respective minima. This is
very important to understand the fate of the Brownian particle's
random walk and its evolution history.
\end{abstract}



\maketitle
%%%%%%%%%%%%%%%%%%%%%%%%%%%%%%%%%%%%%%%%%%%%%%%%%%%%%%%%%%%%%%%%%%%%%%%%%%%%%%
\section{Introduction}
%%%%%%%%%%%%%%%%%%%%%%%%%%%%%%%%%%%%%%%%%%%%%%%%%%%%%%%%%%%%%%%%%%%%%%%%%%%%%%
\label{sec:introduction}

The dynamics of a many-body system can be given in terms of the time
evolution of the probability phase space density of the chosen
ensemble of particles. We must require that the total time derivative
of the probability density vanishes so that the probability density as
seen by an observer moving with a phase point along its phase space
trajectory, is independent of time. Phase points of the statistical
ensemble are neither created nor destroyed as time evolves. 

The Liouvillian dynamics gives rise to the famous kinetic equation
discovered by Boltzmann in 1872 (where the assumption of two body
collisions only and of uncorrelated successive collisions are only
valid at sufficiently low density) or to the exact
Born-Bogoliubov-Green-Kirkwood-Yvon (BBGKY) hierarchical equations. 

According to BBGKY the dynamics of a single particle requires the
knowledge of the two-body probability density. But if we additionally
require that the particle we are looking upon has mass much larger
than that of all the other particles it is possible to expand
perturbatively such dynamics so to find a closed equation for just the
single massive particle probability phase space density.  
  
A colloidal suspension is made of colloidal particles immersed in
a solvent. The solvent (which may be water or other liquids) is in
general a molecular liquid which can be described at a microscopic
level by particles much smaller and lighter than the macromolecular
colloidal particle. 
The problem of the dynamics of the big colloidal solute
particle subject to its interaction with the smaller solvent molecules
in statistical equilibrium at a given absolute temperature $T$ is an
interesting one. Starting from the observation in 1827 by Robert Brown
of the motion of pollen grains in water and the interpretation of
Albert Einstein in 1905 which was later verified experimentally by
Jean Baptiste Perrin in 1908, the Brownian motion began to occupy an
important role in non-equilibrium statistical physics. Even if the
initial description of the colloid solute particle dynamics was an
empirical one it soon 
became clear the microscopic origin of the Brownian motion. A complete
microscopic description has to take into account the solvent in a
proper way. The challenge is to be able to ``remove'' the degrees of
freedom of the solvent in favor of the ones of the solute
particle. So as to have a mesoscopic description of the dynamics of
the colloid solute particle dressed by the solvent. 

In the second half of last century it had been discovered
\cite{Cukier1969} that it is 
indeed possible to derive the Brownian motion equations starting from
the Liouvillian evolution, $\cl\equiv i\{\ch,\ldots\}$, with
$\{\ldots\}$ the Poisson brackets and $\ch$ the model classical
Hamiltonian, of the probability density in the phase 
space of a model of $N$ solvent particles and the colloidal particle,
$f^{[N+1]}(t)=\exp(-i\cl t)f^{[N+1]}(0)$, to the Fokker-Planck
equation \cite{Klein1921,Kramers1940} for the dynamics of the dressed
solute particle alone subject to friction. The friction coefficient
$\xi$ is expressed in terms of an equilibrium average over the phase
space of the solvent in the external field of the solute particle of a
particular term containing the microscopic force exerted by the
solvent on the colloidal particle. At the end of last century it was
furthermore shown that a high friction expansion, in
turn, brings \cite{Bocquet1997} to the Smoluchowski equation for the
colloid solute particle or its equivalent stochastic Langevin equation.

Some related literature to the present framework can be found in the
following references \cite{Balescu,Mazenko,Reichl,Marconi2007}. 

It this work we outline a simple Monte Carlo scheme that may be used
to solve the resulting Smoluchowski equation that we call Dynamical
Monte Carlo (DynMC). The realistic case of a colloidal suspension of
poly(methyl methacrylate) (PMMA) particles in water \cite{Elaissari} 
is taken as an example. In particular we study the approach to the
equilibrium Boltzmann distribution at late times and its resilience
behavior at shorter times as influenced by the steepness of the
external potential and the friction coefficient around their
respective minima. This is very important to understand the fate of
the Brownian particle's random walk and its evolution history.
We test our new algorithm on the case of a harmonic one dimensional
external potential for which the analytic exact solution of the
Smoluchowski equation is known.

Of course a more realistic treatment of the molecular solvent would be
through a quantum statistical mechanics description where $\cl\equiv
i[\hat{\ch},\ldots]/\hbar$, with $[\ldots]$ the commutator and
$\hat{\ch}$ the model Hamiltonian operator. Then we
would have a mixed evolution where the solvent is treated quantum
mechanically and the massive colloid solute particle is treated
classically (see for example Ref. \cite{Schulten1991} or
Ref. \cite{Xu1992} where the full quantum mechanical treatment is
considered). We leave this as an open future problem.

The paper is organized as follows: In Section \ref{sec:l-fp}
we present the reduction from the Liouville equation to the
Fokker-Planck equation, in Section \ref{sec:fp-s} we present the
reduction from the Fokker-Planck equation to the Smoluchowski
equation, in Section \ref{sec:dMC} we present the DynMC algorithm, in
Section \ref{sec:sim} we carry on a computer 
experiment with the newly developed algorithm to solve for the time
dependence of the mean Brownian particle position under the influence
of an external harmonic potential in one spatial solution, Section
\ref{sec:conclusions} is for final remarks.  

%%%%%%%%%%%%%%%%%%%%%%%%%%%%%%%%%%%%%%%%%%%%%%%%%%%%%%%%%%%%%%%%%%%%%%%%%%%%%%
\section{From the Liouville to the Fokker-Planck equation}
%%%%%%%%%%%%%%%%%%%%%%%%%%%%%%%%%%%%%%%%%%%%%%%%%%%%%%%%%%%%%%%%%%%%%%%%%%%%%%
\label{sec:l-fp}

In this section we will reproduce and discuss the multiple time scale
perturbation expansion presented in Ref. \cite{Hansen-McDonald} which
brings from the microscopic Liouville equation for a solute particle
in a solvent to its mesoscopic Fokker-Planck equation. 

We consider a colloidal solute particle of diameter $\Sigma$ and mass
$M$ immersed in a colloidal suspension of small solvent particles of
diameter $\sigma$ and mass $m\ll M$. The Hamiltonian of the system
of $N+1$ particles can thus be written as 
\bq
\ch=\frac{P^2}{2M}+\sum_{i=1}^N\frac{p_i^2}{2m}+V_N(\rr^N)+V_b(\RR,\rr^N),
\eq
where $V_N$ is the total interaction energy of the $N$ solvent
particles of coordinates $\rr^N=(\rr_1,\rr_2,\ldots,\rr_N)$ and
momenta $\pp^N=(\pp_1,\pp_2,\ldots,\pp_N)$, and $V_b$ is the
potential energy of the solvent particles in the field of a Brownian
solute particle placed at $\RR$ with momentum $\PP$. The Liouville
operator splits naturally into a solvent and a Brownian terms:
$\cl=\cl_s+\cl_b$, with 
\bq
\cl_s&=&-i\sum_{i=1}^N\left(\frac{\pp_i}{m}\cdot\frac{\partial}{\partial\rr_i}
+\ff_i\cdot\frac{\partial}{\partial\pp_i}\right),\\
\cl_b&=&-i\left(\frac{\PP}{M}\cdot\frac{\partial}{\partial\RR}
+\FF\cdot\frac{\partial}{\partial\PP}\right),
\eq
where $\ff_i=-\partial(V_N+V_b)/\partial\rr_i$ is the force acting on
the solvent particle $i$ and $\FF=-\partial V_b/\partial\RR$ is the
force exerted on the solute particle by the solvent particles. The
Liouville equation for the phase space probability density of the
system of $N+1$ particles is therefore
\bq \label{liouville}
\frac{\partial}{\partial t}f^{[N+1]}(\BB,\bb^N;t)=
-i(\cl_s+\cl_b)f^{[N+1]}(\BB,\bb^N;t),
\eq
where we we use the following notation: $\BB\equiv\{\RR,\PP\}$ and
$\bb^N\equiv\{\rr^N,\pp^N\}$.

We now introduce a perturbation parameter $\epsilon=\sqrt{m/M}\ll 1$
and rescale the solute particle momenta accordingly as follows:
$\PP'=\epsilon\PP$, so that 
\bq
\cl_b=-i\epsilon\left(\frac{\PP'}{m}\cdot\frac{\partial}{\partial\RR}
+\FF\cdot\frac{\partial}{\partial\PP'}\right)\equiv\epsilon\cl_b'.
\eq
We want to find now the Liouville equation for the one-particle
distribution function
\bq
f_b(\BB,t)\equiv f^{[1]}(\RR,\PP';t)=\int f^{[N+1]}(\BB,\bb^N;t)\,d\bb^N.
\eq
Integrating then Eq. (\ref{liouville}) over the coordinates and
momenta of the solvent we find
\bq \label{liouvilleb}
\frac{\partial}{\partial t}f_b(\BB,t)=-\epsilon\frac{\PP'}{m}\cdot
\frac{\partial}{\partial\RR}f_b(\BB,t)-\epsilon
\int\FF\cdot\frac{\partial}{\partial\PP'}f^{[N+1]}(\BB,\bb^N;t)\,d\bb^N,
\eq
where the term containing $\cl_s$ vanishes because we assume zero net
flow of probability at infinity in phase space.

The perturbation expansion in $\epsilon$ over a {\sl multiple time
  scale} is necessary in order to extract the mesoscopic time
evolution from the microscopic one due to secular divergences of the
solution at sufficiently long times, irrespective of how small
$\epsilon$ may be. We then introduce an auxiliary distribution
function 
\bq \label{pe}
f_\epsilon^{[N+1]}(\BB,\bb^N;t_0,t_1,t_2,\ldots)=
f_{\epsilon 0}^{[N+1]}+\epsilon f_{\epsilon 1}^{[N+1]}+
\epsilon^2 f_{\epsilon 2}^{[N+1]}+\ldots,
\eq
which is a function of multiple time variables
\bq \label{ptl}
t_0=t,~~~t_1=\epsilon t,~~~t_2=\epsilon^2 t,~~~\ldots,~~~
t_n=\epsilon^n t.
\eq
Eq. (\ref{pe}) indicates that the dependence of the distribution
function on $t_n$ characterizes the evolution on the time scale
$t\sim\epsilon^n$ for $n=0,1,2,\ldots$.
So, the original Liouville Eq. (\ref{liouville}) is replaced by
\bq \label{liouvillee}
\left(\frac{\partial}{\partial t_0}+
\epsilon\frac{\partial}{\partial t_1}+
\epsilon^2\frac{\partial}{\partial t_2}+\ldots\right)f^{[N+1]}_\epsilon=
-i(\cl_s+\epsilon\cl_b')f^{[N+1]}_\epsilon,
\eq
and the Liouville equation for the Brownian motion by
\bq \label{liouvillebe}
\left(\frac{\partial}{\partial t_0}+
\epsilon\frac{\partial}{\partial t_1}+
\epsilon^2\frac{\partial}{\partial t_2}+\ldots\right)f_{b\epsilon}=
-\epsilon\frac{\PP'}{m}\cdot\frac{\partial}{\partial\RR}f_{b\epsilon}
-\epsilon\int\FF\cdot\frac{\partial}{\partial\PP'}f_\epsilon^{[N+1]}\,d\bb^N.
\eq
Term by term integration of this equation shows that also
$f_{b\epsilon}$ can be expanded in the form:
$f_{b\epsilon}=f_{b0}+\epsilon f_{b1}+\epsilon^2 f_{b2}+\ldots$. 
The crucial difference between Eq. (\ref{pe}) and a conventional
perturbation expansion is the fact that the auxiliary function has a
physical meaning only along the so-called physical line defined by
(\ref{ptl}). We are therefore free to impose whatever boundary
conditions are needed to ensure that the expansion is free of secular
divergences at successive powers of $\epsilon$ at large times. The
same is true of the expansion for $f_{b\epsilon}$.

We will work up to order $\epsilon^2$, retaining only the three time
variables $t_0$, $t_1$, and $t_2$. Substituting the perturbation
expansion for $f_\epsilon^{[N+1]}$ and $f_{b\epsilon}$ in
Eq. (\ref{liouvillebe}) and equating coefficients of equal powers of
$\epsilon$, we arrive at the following results.
\begin{itemize}
\item[[0.]] To zeroth order in $\epsilon$:\\
From Eq. (\ref{liouvillebe}) we find immediately
\bq \label{0fb0}
\frac{\partial}{\partial t_0}f_{b0}=0,
\eq
so that $f_{b0}=f_{b0}(\RR,\PP';t_1,t_2,\ldots)$. From
Eq. (\ref{liouvillee}) we find
\bq \label{0fe0}
\frac{\partial}{\partial t_0}f_{\epsilon 0}^{[N+1]}=
-i\cl_sf_{\epsilon 0}^{[N+1]}.
\eq

Since the equilibrium phase space probability density of the solvent
in the presence of the Brownian solute particle at $\RR$ satisfies the
relation $\cl_s f_0^{[N]}(\bb^N|\RR)=0$, the solution to
Eq. (\ref{0fe0}) is simply
\bq \label{fe0}
f_{\epsilon 0}^{[N+1]}=f_{b0}(\RR,\PP';t_1,t_2)f_0^{[N]}(\bb^N|\RR),
\eq 
where
\bq
\int f_0^{[N]}(\bb^N|\RR)\,d\bb^N=1,~~~
f_0^{[N]}(\bb^N|\RR)\propto e^{-\beta\ch},~~~
\frac{\partial}{\partial\RR}f_0^{[N]}=\beta\FF f_0^{[N]},
\eq
here $\beta=1/k_BT$ with $T$ the absolute temperature.
We now employ the freedom of choice of boundary
condition on the auxiliary function $f_{b\epsilon}$ imposing the
following initial condition:
$f_{b\epsilon}(\RR,\PP';t_0=0,t_1,t_2)=f_{b0}(\RR,\PP';t_1,t_2)$,
which in turn implies that $f_{bn}(\RR,\PP';t_0=0,t_1,t_2)=0$ for
$n=1,2$.   
\item[[1.]] To first order in $\epsilon$:\\
From Eqs. (\ref{liouvillee}) and (\ref{liouvillebe}) we find
\bq \label{1fe}
\frac{\partial}{\partial t_0}f_{\epsilon 1}^{[N+1]}+
\frac{\partial}{\partial t_1}f_{\epsilon 0}^{[N+1]}=
-i\cl_sf_{\epsilon 1}^{[N+1]}-i\cl_b'f_{\epsilon 0}^{[N+1]}
\eq
and
\bq \label{1fb}
\frac{\partial}{\partial t_0}f_{b1}+
\frac{\partial}{\partial t_1}f_{b0}=
\int(-i\cl_b')f_{\epsilon 0}^{[N+1]}\,d\bb^N,
\eq
respectively. Eqs. (\ref{0fb0}) and (\ref{fe0}) show that $f_{b0}$ and 
$f_{\epsilon 0}^{[N+1]}$ are both independent of $t_0$. To avoid
secular growth of $f_{b1}$ in (\ref{1fb}) it is necessary to impose
$\partial f_{b1}/\partial t_0=0$, which combined with the initial
condition for $f_{b1}$ implies that $f_{b1}=0$ identically, and  
\bq \label{t1fb0}
\frac{\partial}{\partial t_1}f_{b0}=\int(-i\cl_b')f_{\epsilon
  0}^{[N+1]}\,d\bb^N=-i\cl_b'f_{b0},
\eq
where in the last equality we used Eq. (\ref{fe0}). We therefore focus
on the time evolution of $f_{b0}$. Eq. (\ref{t1fb0}) shows that on the
time scale $t_1$ the evolution of the distribution function of the
Brownian particle is the same as that of a single particle in the
field of the solvent. Eq. (\ref{1fe}) can now be rearranged as follows
\bq \nonumber
\left(\frac{\partial}{\partial t_0}+i\cl_s\right)f_{\epsilon
  1}^{[N+1]}&=&
-\left(\frac{\partial}{\partial t_1}+i\cl_b'\right)f_{\epsilon
  0}^{[N+1]}\\ \label{1fe1}
&=&-\FF\cdot\left(\frac{\beta\PP'}{m}+
\frac{\partial}{\partial\PP'}\right)f_{b0}f_0^{[N]},
\eq 
which, upon choosing the boundary condition $f_{\epsilon
  1}^{[N+1]}(\RR,\PP',\bb^N;t_0=0,t_1,t_2)=0$, has the formal solution 
\bq \nonumber
&&f_{\epsilon 1}^{[N+1]}(\RR,\PP',\bb^N;t_0,t_1,t_2)=\\ \label{fe1s}
&&-\int_0^{t_0}ds\,e^{-i\cl_s s}\FF\cdot\left(\frac{\beta\PP'}{m}+
\frac{\partial}{\partial\PP'}\right)f_{b0}(\RR,\PP';t_1,t_2)
f_0^{[N]}(\bb^N|\RR),
\eq
as can be checked by direct substitution in (\ref{fe1s}).
\item[[2.]] To second order in $\epsilon$:\\
From Eq. (\ref{liouvillebe}) and the fact that $f_{b1}=0$ we find
\bq \label{2fb}
\frac{\partial}{\partial t_0}f_{b2}+
\frac{\partial}{\partial t_2}f_{b0}=
\int(-i\cl_b')f_{\epsilon 1}^{[N+1]}\,d\bb^N.
\eq
Since $f_{b0}$ is independent of $t_0$, secular growth is again
suppressed by setting $\partial f_{b2}/\partial t_0=0$. Substituting
the solution (\ref{fe1s}) into Eq. (\ref{2fb}) we obtain a closed
equation for the evolution of $f_{b0}(\RR,\PP';t_1,t_2)$:
\bq \nonumber
\frac{\partial}{\partial t_2}f_{b0}&=&\lim_{t_0\to\infty}\int d\bb^N\,
f_0^{[N]}(\bb^N|\RR)i\cl_b'\\
&&\times\int_0^{t_0}ds\,e^{-i\cl_s s}\FF\cdot\left(\frac{\beta\PP'}{m}+
\frac{\partial}{\partial\PP'}\right)f_{b0},
\eq
where the limit $t_0\to\infty$ can be easily taken because $f_{b0}$ is
independent of $t_0$. Recognizing that $e^{-i\cl_s s}\FF$ is the time
evolution $\FF(-s)$ we then find
\bq \nonumber
\frac{\partial}{\partial t_2}f_{b0}&=&\lim_{t_0\to\infty}
\int_0^{t_0}ds\,\langle\cl_b'\FF(-s)\rangle_b\cdot\left(
\frac{\beta\PP'}{m}+\frac{\partial}{\partial\PP'}\right)f_{b0}\\
&=&\frac{1}{3}\int_0^\infty ds\,\langle\FF\cdot\FF(-s)\rangle_b
\frac{\partial}{\partial\PP'}\cdot\left(
\frac{\beta\PP'}{m}+\frac{\partial}{\partial\PP'}\right)f_{b0},
\eq
where $\langle\ldots\rangle_b=\int\ldots f_0^{[N]}(\bb^N|\RR)\,d\bb^N$
denotes an equilibrium average over the phase space variables of the
solvent particles in the external field of the solute particle. 

Putting together the results obtained so far and going back to
physical time and to the original momentum variable $\PP$, we arrive
finally to the Fokker-Planck (or Klein \cite{Klein1921} and Kramers
\cite{Kramers1940}) equation for $f_b(\RR,\PP;t)$: 
\bq \nonumber
\frac{\partial}{\partial t}f_b(\RR,\PP;t)&=&
\left.\left(\epsilon\frac{\partial}{\partial t_1}+
\epsilon^2\frac{\partial}{\partial t_2}\right)
\right|_{t_1=\epsilon t,t_2=\epsilon^2 t}f_{b0}(\RR,\PP;t_1,t_2)\\ \label{FP}
&=&\left[-\frac{\PP}{M}\cdot\frac{\partial}{\partial\RR}+
\xi\frac{\partial}{\partial\PP}\cdot\left(\PP+\frac{M}{\beta}
\frac{\partial}{\partial\PP}\right)\right]f_b(\RR,\PP;t),
\eq
where the {\sl friction coefficient}
\bq
\xi=\frac{\beta}{3M}\int\langle\FF\cdot\FF(-s)\rangle_b\,ds.
\eq
\end{itemize}

%%%%%%%%%%%%%%%%%%%%%%%%%%%%%%%%%%%%%%%%%%%%%%%%%%%%%%%%%%%%%%%%%%%%%%%%%%%%%%
\section{From the Fokker-Planck to the Smoluchowski equation}
%%%%%%%%%%%%%%%%%%%%%%%%%%%%%%%%%%%%%%%%%%%%%%%%%%%%%%%%%%%%%%%%%%%%%%%%%%%%%%
\label{sec:fp-s}

We now introduce a dimensionless friction $\xi_d=\xi\tau_T$ with
$\tau_T=\Sigma/v_T$ a thermal time and $v_T=\sqrt{1/\beta M}$ the
thermal velocity. 
In the high friction limit one can carry on a multiple time scale
perturbation expansion in the small $\epsilon_b\equiv 1/\xi_d\ll 1$
parameter \cite{Bocquet1997}, which  multiplies the time derivative in
the Fokker-Planck equation (\ref{FP}), to finally reach the
Smoluchowski \cite{Smoluchowski1916} equation, governing the time
evolution of the probability density in configuration space:
\bq \label{rhob}
\rho_b(\RR,t)=\int f_b(\RR,\PP;t)\,d\PP.
\eq
We then repeat the same multiple time scale perturbation analysis
performed in the previous section replacing now $\epsilon$ with
$\epsilon_b$. If we use the following notation
\bq
\cl_\text{FP}\equiv\frac{\partial}{\partial\PP}\cdot
\left(\PP+\frac{M}{\beta}\frac{\partial}{\partial\PP}\right),
\eq
the identification of different powers of $\epsilon_b$ in the
Fokker-Planck equation gives the following relations:
\bq
\cl_\text{FP}f_{b0}&=&0,\\
\cl_\text{FP}f_{b1}&=&\tau_T\left[\frac{\partial}{\partial t_0}+
\frac{\PP}{M}\cdot\frac{\partial}{\partial\RR}\right]f_{b0},\\
\cl_\text{FP}f_{b2}&=&\tau_T\left[\frac{\partial}{\partial t_0}+
\frac{\PP}{M}\cdot\frac{\partial}{\partial\RR}\right]f_{b1}+
\tau_T\frac{\partial}{\partial t_1}f_{b0}.
\eq

The zeroth order equation imposes a Maxwellian distribution
\bq
f_{b0}(\RR,\PP;t_0,t_1,\ldots)=\Phi(\RR;t_0,t_1,\ldots)
e^{-\beta P^2/2M}.
\eq

The first order equation imposes then
\bq
\cl_\text{FP}f_{b1}=\tau_T\frac{\partial\Phi}{\partial t_0}e^{-\beta
  P^2/2M}+\tau_T\frac{\PP}{M}\cdot\frac{\partial\Phi}{\partial\RR}
e^{-\beta P^2/2M}.
\eq
In order to eliminate secular divergences we must require
$\partial\Phi/\partial t_0=0$, and the first correction for the
distribution function is now given by 
\bq
f_{b1}(\RR,\PP;t_0,t_1,\ldots)=-\tau_T\frac{\PP}{M}\cdot
\frac{\partial\Phi}{\partial\RR}e^{-\beta P^2/2M}+
\Psi(\RR;t_0,t_1,\ldots)e^{-\beta P^2/2M}.
\eq

The second order equation becomes 
\bq \nonumber
\cl_\text{FP}f_{b2}&=&\left[\tau_T\frac{\partial\Psi}{\partial t_0}+
\tau_T\frac{\partial\Phi}{\partial t_1}-(v_T\tau_T)^2
\frac{\partial}{\partial\RR}\cdot\frac{\partial\Phi}{\partial\RR}
\right]e^{-\beta P^2/2M}+\tau_T\frac{\PP}{M}\cdot
\frac{\partial\Psi}{\partial\RR}e^{-\beta P^2/2M}+\\ \label{fp2f}
&&\left[\tau_T^2\left(v_T^2\frac{\partial}{\partial\RR}\cdot
\frac{\partial}{\partial\RR}
-\frac{\PP}{M}\cdot\frac{\partial}{\partial\RR}
\frac{\PP}{M}\cdot\frac{\partial}{\partial\RR}\right)\Phi\right]
e^{-\beta P^2/2M}.
\eq
In order to remove the secular divergences as $t_0$ grows to infinity,
we must impose $\partial\Psi/\partial t_0=0$. Moreover it is easy to
show that the eigenfunctions of $\cl_\text{FP}$ are the functions
$H_n$ $\exp(-\beta P^2/2M)$ where $H_n$ is the $n$th Hermite
polynomial. Since the Maxwellian is associated with a null eigenvalue,
in order to require that the integral of the right hand side of
Eq. (\ref{fp2f}) in $d\PP$ vanishes, we must impose that all terms
multiplying the Maxwellian vanish, due to the orthogonality condition
between the eigenfunctions. We must then have    
\bq \label{1Phi}
\frac{\partial\Phi}{\partial t_1}=
v_T^2\tau_T\frac{\partial}{\partial\RR}\cdot
\frac{\partial\Phi}{\partial\RR}. 
\eq
We now can collect the results obtained so far to obtain the
distribution function to order $\epsilon_b^2$ as follows
\bq
f_b(\RR,\PP;t)=\left[\Phi-\epsilon_b\tau_T
\frac{\PP}{M}\cdot\frac{\partial\Phi}{\partial\RR}+
\epsilon_b\Psi+O(\epsilon_b^2)\right]e^{-\beta P^2/2M},
\eq
and from Eq. (\ref{rhob}) it follows immediately
\bq
\rho_b(\RR,t)=\left(\frac{2M\pi}{\beta}\right)^{3N/2}
\left[\Phi+\epsilon_b\Psi+O(\epsilon_b^2)\right].
\eq
Then the time evolution for $\rho_b$ is obtained restricting the
different variables $t_i$ to the physical line as follows
\bq \nonumber
&&\frac{\partial}{\partial t}\rho_b(\RR,t)\\ \nonumber
&=&\left(\frac{\partial}{\partial t_0}+
\epsilon_b\frac{\partial}{\partial t_1}\right)
\left(\frac{2M\pi}{\beta}\right)^{3N/2}
[\Phi(\RR;t_1,\ldots)+\epsilon_b\Psi(\RR;t_1,\ldots)+
O(\epsilon_b^2)]_{t_0=t,t_1=\epsilon_b t, \ldots}\\
&=&\epsilon_b\frac{\partial}{\partial t_1}
\left(\frac{2M\pi}{\beta}\right)^{3N/2}\left.\Phi(\RR;t_1,\ldots)
\right|_{t_0=t,t_1=\epsilon_b t, \ldots}+O(\epsilon_b^2).
\eq
Then using Eq. (\ref{1Phi}) we finally find the diffusion
equation 
\bq
\frac{\partial\rho_b(\RR,t)}{\partial t}=
\frac{v_T^2}{\xi}\frac{\partial}{\partial\RR}\cdot
\frac{\partial\rho_b(\RR,t)}{\partial\RR}+O(1/\xi_d^2).
\eq

If we had an external field
$\FF_e(\RR)=-\partial V_e(\RR)/\partial\RR$ acting on the Brownian
solute particle then the initial Hamiltonian would have been
\bq
\ch=\frac{P^2}{2M}+\sum_{i=1}^N\frac{p_i^2}{2m}+V_N(\rr^N)+V_b(\RR,\rr^N)
+V_e(\RR),
\eq
and the Smoluchowski equation \cite{Smoluchowski1916} becomes
\bq \label{smol}
\frac{\partial\rho_b(\RR,t)}{\partial t}=
\frac{v_T^2}{\xi}\frac{\partial}{\partial\RR}\cdot\left(
\frac{\partial}{\partial\RR}-\beta\FF_e(\RR)\right)\rho_b(\RR,t),
\eq
which can also be written as 
\bq
\frac{\partial\rho_b(\RR,t)}{\partial t}=
\frac{v_T^2}{\xi}\frac{\partial}{\partial\RR}e^{-\beta V_e(\RR)}
\frac{\partial}{\partial\RR}e^{\beta V_e(\RR)}\rho_b(\RR,t),
\eq
which shows immediately that $\rho_b\propto\exp[-\beta V_e(\RR)]$ is a
stationary solution. So that calling 
\bq \label{fdt}
\beta v_T^2/\xi=1/\gamma,
\eq
where $\gamma$ is usually known as the {\sl mobility}, we can rewrite
the Smoluchowski equation (\ref{smol}) as  
\bq
\frac{\partial\rho_b(\RR,t)}{\partial t}=
\frac{\partial}{\partial\RR}\cdot\left(
\frac{\partial}{\partial\RR}\frac{v_T^2}{\xi}-
\frac{\FF_e(\RR)}{\gamma}\right)\rho_b(\RR,t),
\eq
which trough Ito's calculus \cite{Gardiner1983} can be proved to be
equivalent to the following stochastic differential equation, the
Langevin equation 
\bq \label{langevin}
\gamma\dot{\RR}=\FF_e(\RR)+\sqrt{2(\gamma v_T)^2/\xi}\,\zeta(t),
\eq
where we denote with the dot a time derivative and $\zeta$ is a white
noise. In this respect Eq. (\ref{fdt}) is considered as a case of
{\sl fluctuation and dissipation theorem}. It is then straightforward
to show that for $\FF_e=0$ we find
$\langle[\RR(t)-\RR(0)]^2\rangle=6Dt$ with $D=v_T^2/\xi$ the {\sl
  diffusion constant}. Equation (\ref{fdt}) is known as the Einstein
relation. The fluctuations of the particle as it undergoes its random
walk are related to the drag force (or dissipation of momentum) that
the particle feels as it moves through the solvent. 

The Einstein relation gives an excellent way to determine Boltzmann's
constant experimentally. Watch a particle perform a Brownian
jitter. After time $t$, the distance traveled by the particle should
be $\langle R^2\rangle=2k_BTt/\pi\eta\Sigma$, where we have used the
Stokes formula $\gamma=3\pi\eta\Sigma$ to relate the mobility to the
viscosity $\eta$. This experiment was done in 1908 by the French
physicist Jean Baptiste Perrin and won him the 1926 Nobel prize.

The full form of the Langevin equation is the one with the
acceleration term. Then using the relation $\PP/M=\VV=\dot{\RR}$ we
should have in a less viscous fluid the equation
$M\dot{\VV}=-\gamma\VV+\FF_e+\gamma\sqrt{2D}\zeta$ for which we still
find for $\FF=0$, $\langle[\RR(t)-\RR(0)]^2\rangle=6Dt$ and 
$\langle\VV(t)\cdot\VV(t)\rangle=3D\gamma/M$ at late 
times. This stochastic differential equation leads through Ito's
calculus to the Fokker-Planck equation for $f_b(\RR,\PP;t)$. This
equation (\ref{FP}) is sometimes also called the Klein and Kramers
equation and sometimes the Chandrasekhar equation.
 
%%%%%%%%%%%%%%%%%%%%%%%%%%%%%%%%%%%%%%%%%%%%%%%%%%%%%%%%%%%%%%%%%%%%%%%%%%%%%%
\section{The dynamical Monte Carlo method}
%%%%%%%%%%%%%%%%%%%%%%%%%%%%%%%%%%%%%%%%%%%%%%%%%%%%%%%%%%%%%%%%%%%%%%%%%%%%%%
\label{sec:dMC}

We now want to show how the Smoluchowski evolution equation can be
solved by stochastic means. In nature, equilibrium distributions are
generated by an evolution process. The Smoluchowski equation  
\bq \label{smo}
\frac{\partial\rho_b(\RR,t)}{\partial t}=\bnabla\cdot\DD(\RR)\cdot
[\bnabla-\beta \FF_e(\RR)]\rho_b(\RR,t),
\eq
is the unique ``master'' equation which is:
\begin{itemize}
\item local in space
\item goes to the Boltzmann distribution
\item is Markovian
\end{itemize}
Here $\DD(\RR)$ is, in general, a many-body tensor. In the previous
section we saw that for the one-body problem it is sufficient to take
it as a constant diagonal tensor $D_{ij}=\delta_{ij}D$ and
$\FF_e=-\bnabla V_e$ is the external force.  

The asymptotic solution of $\rho_b(\RR,t)$ will be $\rho_0(\RR)\propto
\exp[-\beta V_e(\RR)]$. It is easy to see that this distribution
satisfies $d\rho_0/dt=0$. If we assume the process is ergodic, since
it is Markovian, this must be the only solution.

Let us define Green's function: $G(\RR,\RR_0;t)=G(\RR_0\to\RR;t)$
is the solution to Eq. (\ref{smo}) with the boundary condition at zero
time: $G(\RR,\RR_0;0)=\delta(\RR-\RR_0)$. We can prove that the Green's
function satisfies {\sl detailed balance}: 
\bq
\rho_0(\RR)G(\RR\to\RR';t) =\rho_0(\RR')G(\RR'\to\RR;t),
\eq
for any value of $t$. To do that one writes the evolution equation
for the symmetrized Green's function:
$[\rho_0(\RR)/\rho_0(\RR')]^{1/2}$ $G(\RR\to\RR';t)$, and sees that
the right hand side of the master equation is a Hermitian operator
(see Appendix \ref{app:1}) which implies that the symmetrized Green's
function is symmetric in $\RR$ and $\RR'$. Then the random walk must
eventually have $\rho_0$ as its equilibrium distribution
\cite{Kalos-Whitlock}.  
$G$ can be used for a transition probability and it will always give
an acceptance probability of unity \cite{Metropolis1953}. Also it
gives the interesting dynamics of a viscous particles always in
contact with a heat bath. 

The Smoluchowski equation leads to an interesting process but we can
only calculate $G$ in the short time limit. In the following we explain
a general procedure for devising an algorithm of sampling $G$. Let us
calculate the moments of $G$, 
\bq
I_n(\RR_0,t)=\int d\RR\,(\RR-\RR_0)^n G(\RR_0\to\RR;t).
\eq 
Take the time derivative of this equation, use the master equation on
the right hand side, and integration by parts to get a simple integral
over $G$ on the right hand side. We interpret this as an average
$\langle\ldots\rangle$. We assume there are no absorbing surfaces of
the random walks. Then, 
\bq
dI_0/dt=0.
\eq
This implies the normalization of $G$ is always one, so the evolution
describes a process which neither creates nor destroys walks. The next
moment is: 
\bq
d[I_1]_k/dt=\langle\bnabla_jD_{kj}+D_{kj}[F_e]_j\beta\rangle,
\eq
where we use the Einstein summation convention over the repeated
indexes. Let us assume that $\FF_e$ and $\bnabla\DD$ are slowly
varying. Then we can replace them by the values at the initial point
and integrate in time: 
\bq \label{averageR}
[R_t]_k=\langle R_k\rangle=[R_0]_k+
t\langle\bnabla_jD_{kj}(\RR)+D_{kj}(\RR)[F_e]_j(\RR)\beta\rangle_g|_{\RR=\RR_0}
+O(t^2), 
\eq
where $\langle\ldots\rangle_g$ is an average respect to the small time
Green's function $G_g$ of Eq. (\ref{Gg}). The equation for the second
moment is: 
\bq \label{I2}
dI_2/dt=2\langle D_{ii}+(R-R_0)_i(\bnabla_jD_{ij}+D_{ij}[F_e]_j\beta)\rangle,
\eq
Integrating in time we then find
\bq \label{sigmaR}
\langle(\RR-\RR_0)^2\rangle=2\langle D_{ii}(\RR)\rangle_g|_{\RR=\RR_0}t+O(t^2).
\eq
The solution at small time is a Gaussian distribution with the above
mean and covariance, namely
\bq \label{Gg} 
G_g(\RR,\RR_0;t)=\frac{e^{-(\RR-\RR_t)
(4\langle D_{ii}(\RR)\rangle_g|_{\RR=\RR_0}t)^{-1}(\RR-\RR_t)}}
{\sqrt{4\pi\langle D_{ii}(\RR)\rangle_g|_{\RR=\RR_0}t}^{3N}}.  
\eq
According to the central limit theorem,
Eqs. (\ref{averageR}) and (\ref{sigmaR}) are all that is needed to
simulate the random walk if the time step $t$ is sufficiently small.
The effect of the external field is to push the mean position of the
Brownian particle away from its current position. An outward push in
directions where the external force is positive. The cage of the
surrounding solvent is also present whenever the diffusion tensor is
$\RR$ dependent.  

%%%%%%%%%%%%%%%%%%%%%%%%%%%%%%%%%%%%%%%%%%%%%%%%%%%%%%%%%%%%%%%%%%%%%%%%%%%%%%
\section{A simple Monte Carlo simulation}
%%%%%%%%%%%%%%%%%%%%%%%%%%%%%%%%%%%%%%%%%%%%%%%%%%%%%%%%%%%%%%%%%%%%%%%%%%%%%%
\label{sec:sim}

For some simple forms of the external potential $V_e$ the Smoluchowski
equation admits an analytical solution. For example the
one-dimensional Smoluchowski equation in empty space (which requires
the probability density to vanish at spatial infinity faster than any
inverse power of $R$) can be solved
analytically for
\begin{itemize}
\item A {\sl linear} potential $V_e(x)=c x$ \cite{Kubo1969},
\bq
G(x,x_0;t)=\frac{e^{-\frac{(x-x_0+D\beta ct)^2}{4Dt}}}{\sqrt{4\pi Dt}}.
\eq
\item A {\sl harmonic} potential $V_e(x)=c x^2/2$ \cite{Sun2005},
\bq
G(x,x_0;t)&=&\frac{e^{-\frac{\left(x-x_0e^{-2t/\tilde{t}}\right)^2}{2S(t)/\beta
      c}}}{\sqrt{2\pi S(t)/\beta c}},\\ \nonumber
S(t)&=&1-e^{-4t/\tilde{t}},\\ \nonumber
\tilde{t}&=&2/\beta cD.
\eq
\end{itemize}
In each one of these cases a proper change of time dependent coordinates
is used to reduce the Smoluchowski equation to the free diffusion
equation.

Except for these simple cases in order to extract the evolution of the
Green's function we must resort to a simulation. We can then use the
Markovian property as follows
\bq \nonumber
&&G(\RR,\RR_0;t-t_0)=\\ \label{green}
&&\int G_g(\RR,\RR_n;t-t_n)G_g(\RR_n,\RR_{n-1};t_n-t_{n-1})
\cdots G_g(\RR_1,\RR_0;t_1-t_0)\,d\RR_1d\RR_2\cdots d\RR_n,
\eq
where $t_n-t_{n-1}=\tau$ is a small time step. At fixed $\RR, \RR_0$,
and $t-t_0=(n+1)\tau$ we can then easily compute the Green's function
$G(\RR,\RR_0;t)$ through a Monte Carlo integration sampling the
successive $G_g$ of Eq. (\ref{Gg}) with the Box-M\"uller method (see
Ref. \cite{Kalos-Whitlock} section 3.1). For small enough $\tau$ we
sample $G_g(\RR_1,\RR_0;\tau)$ at fixed $\RR_0$ generating $\RR_1$
then we sample $G_g(\RR_2,\RR_1;\tau)$ generating $\RR_2$ and so on
for the remaining $n-2$ factors ending with the generated $\RR_n$. We
can then evaluate $G_g(\RR,\RR_n;\tau)$ at fixed $\RR$ on $\RR_n$ with
a Monte Carlo integration to find $G(\RR,\RR_0;(n+1)\tau)$. Clearly we
will have to control the convergence of the algorithm as $\tau$ gets
smaller and smaller.

Eq. (\ref{green}) suggests a path integral representation of the
Green function solution of the Smoluchowski equation, namely
\bq
G(\RR,\RR_0;t-t_0)\propto\pint_{\RR_0}^\RR e^{-\cs}\,\cd\RR(t),
\eq
with an action
\bq 
\cs=\int_{t_0}^tdt'\,\frac{\sum_k\left[\dot{R}_k(t')-
\langle\nabla_jD_{kj}+D_{kj}[F_e]_j\beta\rangle|_{\RR=\RR(t')}\right]^2}{
4\langle D_{ii}\rangle|_{\RR=\RR(t')}},
\eq
where the dot denotes a total derivative with respect to time.

If the colloidal particle is initially localized around
$\bar{\RR}_0=\int\RR\rho_{b0}(\RR)\,d\RR$ where we assume to be given 
$\rho_{b0}(\RR)=\rho_b(\RR,t_0)$ we may want to find its average
position at a later time
\bq
\bar{\RR}(t)=\int\RR G(\RR,\RR_0;t-t_0)\rho_{b0}(\RR_0)\,d\RR d\RR_0.
\eq

\begin{figure}[htbp]
\begin{center}
\includegraphics[width=10cm]{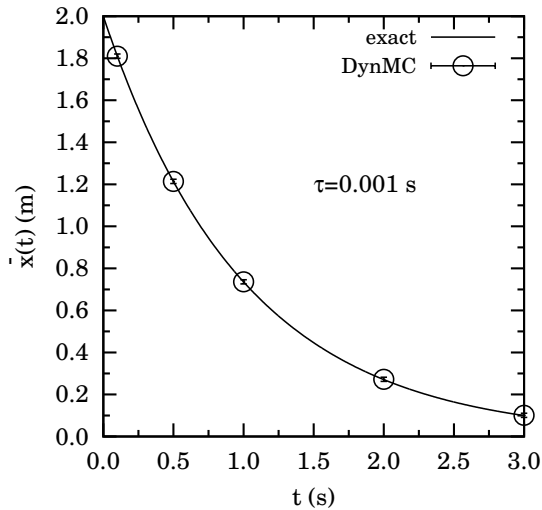}
\end{center}  
\caption{DynMC simulation for the one dimensional harmonic potential
  with an initial gaussian distribution centered on $x_0=2$~m with
  $\beta cD=1~\text{s}^{-1}$ and $\tau=0.001$~s. In this case $\langle
  F_e(x)\rangle_g=-cx/(1+\beta cD\tau)$. The continuous line is the
  exact analytical solution of Eq. (\ref{hp1dax}) and the symbols are
  the results of the simulation.} 
\label{fig:DynMC}
\end{figure}

Relevant mesoscopic time scales are the time $\tau_b=1/\xi$ over which
the velocity of a Brownian particle relaxes. At room temperature, for
typical values of $\xi$, we find $\tau_b\approx 10^{-9}$~s. Taking
$\Sigma\approx 1\mu m$ and $M\approx 10^{-21}$~kg
\cite{Elaissari,Klein2003} at 
room temperature $1/\beta\approx 10^{-21}$~J we have $v_T\approx 1$~m/s
and $\tau_T\approx 10^{-6}$~s. Another relevant time scale is the time
required for an isolated Brownian particle to diffuse over a distance
equal to its diameter $\tau_c=\Sigma^2/D=\xi \beta M\Sigma^2\approx
10^{-3}~\text{s}>\tau_T>\tau_b$. On the microscopic level the shortest time
scale is the Enskog mean collision time
$\tau_E=1/\Gamma_E=\sqrt{\beta m/\pi}/4\rho_s\sigma^2g(\sigma)$ where
$g(r)$ (see Ref. \cite{Hansen-McDonald} section 2.5)  is the radial
distribution function of the solvent. For a solvent such as water
$\sigma=\Sigma/10^4$, and $\tau_E\approx 10^{-12}$~s. We already know
that for $t\gg\tau_E$ the Smoluchowski solution tends to the
equilibrium Boltzmann distribution which remains then stable on a
large subsequent time scale. For example, for the one dimensional
harmonic potential with an initial gaussian distribution centered on
$x_0$ we find 
\bq \label{hp1dax}
\bar{x}(t)=\frac{\sqrt 2 x_0e^{-\beta cDt}}{\sqrt{(1-e^{-2\beta cDt})
[1+\coth(\beta cDt)]}},
\eq
showing that the relaxation time is proportional to $1/\beta cD$.
In Fig. \ref{fig:DynMC} we show a DynMC simulation of this case for
$\beta cD=1~\text{s}^{-1}, x_0=2$~m, and fixed $\tau=0.001$~s.
Clearly at long enough times the solute particle will reach the
equilibrium average position
\bq
\bar{\RR}_e=\frac{\int \RR e^{-\beta V_e(\RR)}\,d\RR}
{\int e^{-\beta V_e(\RR)}\,d\RR}.
\eq
The natural time scale over which the Smoluchowski evolution approach
equilibrium is $\tau_b$. What may be interesting to observe is whether
a proper control of the external fields allows the approach to
equilibrium to happens more rapidly, thus indicating a ``persistence''
of the Boltzmann distribution since early times. Clearly if
$V_e(\RR)=V_e(R)$ one must end up in $\bar{\RR}_e=0$, by symmetry. For
a single Brownian particle Eq. (\ref{averageR}) becomes 
\bq
\RR_\tau\simeq\RR_0+\tau D\beta\langle\FF_e\rangle_g|_{\RR=\RR_0},
\eq
so that, if $\langle\FF_e\rangle_g=0$ we will have $\RR_t=\RR_0$ at all
time slices and
\bq
\bar{\RR}(t)=\int\RR G_g(\RR,\RR_0;t-t_0)\rho_{b0}(\RR_0)\,d\RR d\RR_0.
\eq
We also immediately see that the persistence will be the more
important the steeper the external potential approaches its
minimum. On an unstable equilibrium point of the potential (a maximum)
the Green's function will be swallowed and eventually vanish in time.

Another interesting question is to determine the influence on the
evolution of a spatially dependent diffusion coefficient
$D(\RR)=v_T^2/\xi(\RR)$ rather than a constant. From
Eq. (\ref{averageR}) we see that around an equilibrium point for the
external potential we will still have
\bq
\RR_\tau\simeq\RR_0+\tau\langle\bnabla D\rangle_g|_{\RR=\RR_0}.
\eq
We will call this the ``cage effect'' of the solvent on the evolution
of the solute particle.

In addition to the empty space boundary conditions other kinds are
possible. Calling $\ci\equiv D(\bnabla-\beta\FF_e)$ the flux operator
and $\partial\Omega$ the boundary of the space $\Omega$ where the
solute particle is confined, the most general case are the {\sl
  radiation boundary conditions}, namely
\bq
\hat{\nn}(\RR)\cdot\ci(\RR)G(\RR,\RR_0;t-t_0)=\omega
G(\RR,\RR_0;t-t_0),~~~\RR\in\partial\Omega,
\eq
where $\hat{\nn}(\RR)$ denotes a unit vector normal to the surface
$\partial\Omega$. The reactivity is measured by the $\omega$
parameter. For $\omega=0$ we have a non-reactive or {\sl reflective
  boundary condition} and for $\omega\to\infty$ we have a {\sl
  reaction boundary condition}. So for a confined case we have to
choose the required boundary condition and then determine the proper
small time step Green's function $G_g$. For example for the solution
in half space $X\ge 0$ with a reaction boundary condition at $X=0$,
$G((0,Y,Z),\RR_0;t-t_0)=0$, we may use the image technique as follow 
\bq \nonumber
G_g(\RR,\RR_0;\tau)&=&\frac{e^{-[(X-X_t)^2+(Y-Y_t)^2+(Z-Z_t)^2]/
4\langle D_{ii}(\RR)\rangle_g|_{\RR=\RR_0}\tau}}
{\sqrt{4\pi\langle D_{ii}(\RR)\rangle_g|_{\RR=\RR_0}t}^{3N}}-\\
&&\frac{e^{-[(X+X_t)^2+(Y-Y_t)^2+(Z-Z_t)^2]/4\langle D_{ii}(\RR)\rangle_g|_{\RR=\RR_0}\tau}}
{\sqrt{4\pi\langle D_{ii}(\RR)\rangle_g|_{\RR=\RR_0}\tau}^{3N}}.
\eq   

%%%%%%%%%%%%%%%%%%%%%%%%%%%%%%%%%%%%%%%%%%%%%%%%%%%%%%%%%%%%%%%%%%%%%%%%%%%%%%
\section{Conclusions}
%%%%%%%%%%%%%%%%%%%%%%%%%%%%%%%%%%%%%%%%%%%%%%%%%%%%%%%%%%%%%%%%%%%%%%%%%%%%%%
\label{sec:conclusions}

In conclusion, for a colloidal suspension made up of one solute big and
heavy particle and $N$ solvent small and light particles 
it is possible to reduce the Liouville dynamics to the Fokker-Planck
dynamics by integrating over the coordinates and momenta of the solvent
particles. Integrating furthermore on the momenta of the
solute particle it is possible to reduce the Fokker-Planck equation to
the Smoluchowski equation in the high friction regime. The two
successive reductions are based each on a multiple time scale perturbation
expansions. The first reduction was discovered in the second half of
last century and the second reduction at the end of last century.

A simple Dynamical Monte Carlo (DynMC) algorithm is
presented to solve the Smoluchowski equation for the Brownian motion
of the solute particle in empty space and a thought computer
experiment illustrated.
We discuss how a steep external potential around its minimum will
drive the approach to equilibrium more rapidly than the natural time
scale $\tau_b$ and the cage effect due to the solvent. In particular
we present a DynMC simulation on the harmonic case, with a constant
diffusion, which well reproduces the exact analytical solution of the
Smoluchowski equation. The simplicity of the simulation makes it very
useful for the treatment of problems which do not have an analytic
solution. We presented results for the average position of the solute
particle as a function of time, but other quntities can be computed as
for example the full Green's function starting from Eq. (\ref{green}). 

We discuss the proper boundary conditions to be imposed on the
Smoluchowski second order partial differential equation in order to
have a radiation, reflective, or reaction boundary and we give the
small time expression for the Green's function in half space with a
reaction boundary condition on the plane of separation using the image
technique. 

An interesting progress that could be made at the level of the
Fokker-Planck description is to derive the expression for the friction
coefficient for a colloidal patchy sticky hard sphere in a solvent of
isotropic sticky hard spheres, in the spirit of
Refs. \cite{Fantoni05a,Fantoni05b,Fantoni06a,Fantoni06b,Fantoni06c,Fantoni07,Fantoni08a,Fantoni09b,Fantoni11a,Fantoni12a,Fantoni12c,Fantoni13c,Fantoni13e,Fantoni13f,Fantoni13h,Fantoni15a,Fantoni15c} or in a
solvent of penetrable square well particles, in the spirit of
Refs. \cite{Fantoni08b,Fantoni09a,Fantoni10a,Fantoni10b,Fantoni11b,Fantoni11d}
or fluid mixtures adsorbed in porous disordered materials, as in
Ref. \cite{Lucentini08,Pellicane08}.

We leave as an open problem the generalization of the present
treatment to the case of a classical Brownian solute macromulecule in
a quantum solvent.

\appendix
%%%%%%%%%%%%%%%%%%%%%%%%%%%%%%%%%%%%%%%%%%%%%%%%%%%%%%%%%%%%%%%%%%%%%%%%%%%%%%
\section{Hermiticity of the Smoluchowski operator} 
%%%%%%%%%%%%%%%%%%%%%%%%%%%%%%%%%%%%%%%%%%%%%%%%%%%%%%%%%%%%%%%%%%%%%%%%%%%%%%
\label{app:1} 

We will work in empty space. Then we introduce the Smoluchowski
operator $\cl_S(\RR)=\bnabla\DD\rho_0(\RR)\bnabla\rho_0^{-1}(\RR)$. 
For any two functions $g(\RR)$ and $h(\RR)$ decaying to zero at
spatial infinity faster than any inverse power we can write, using two
times an integration by parts,
\bq \nonumber
&&\int d\RR\,\frac{1}{\sqrt{\rho_0(\RR)}}g(\RR)\cl_S(\RR)
\sqrt{\rho_0(\RR)}h(\RR)=\\ \nonumber
&&-\int d\RR\,\rho_0(\RR)\bnabla\left(\frac{1}{\sqrt{\rho_0(\RR)}}
g(\RR)\right)\DD\bnabla\frac{1}{\sqrt{\rho_0(\RR)}}h(\RR)=\\
&&\int d\RR\,\frac{1}{\sqrt{\rho_0(\RR)}}h(\RR)\cl_S(\RR)
\sqrt{\rho_0(\RR)}g(\RR),
\eq
which proves the Hermiticity,
$[\sqrt{\rho_0}^{-1}\cl_S\sqrt{\rho_0}]^\dagger=
\sqrt{\rho_0}^{-1}\cl_S\sqrt{\rho_0}$, of the Smoluchowski
operator. So that, from the {\sl Chapman-Kolmogorov} relation
\bq
G(\RR_0\to\RR;t)=\int d\RR_1\,G(\RR_1\to\RR;\tau)G(\RR_0\to\RR_1;t-\tau),
\eq 
follows
\bq \nonumber
\frac{\partial}{\partial t}G(\RR_0\to\RR;t)&=&
\cl_S(\RR)G(\RR_0\to\RR;t)\\ \nonumber
&=&\int d\RR_1G(\RR_1\to\RR;\tau)\cl_S(\RR_1)G(\RR_0\to\RR_1;t-\tau)
\\ \nonumber
&=&\int d\RR_1\frac{G(\RR_0\to\RR_1;t-\tau)}{\rho_0(\RR_1)}
\cl_S(\RR_1)\rho_0(\RR_1)G(\RR_1\to\RR;\tau)\\ \label{man}
&\stackrel{\tau\to t}{\longrightarrow}&
\frac{1}{\rho_0(\RR_0)}\cl_S(\RR_0)\rho_0(\RR_0)G(\RR_0\to\RR;t),
\eq
where in the last limit we used the initial condition
$G(\RR_0\to\RR_1;0)=\delta(\RR_0-\RR_1)$. And, using the symmetry of
the left hand side of Eq. (\ref{man}) respect to exchange of
$\RR_0\leftrightarrow\RR$, we finally find the detailed balance
relation $\rho_0(\RR_0) G(\RR_0\to\RR;t)=\rho_0(\RR) G(\RR\to\RR_0;t)$.
 
%%%%%%%%%%%%%%%%%%%%%%%%%%%%%%%%%%%%%%%%%%%%%%%%%%%%%%%%%%%%%%%%%%%%%%%%%%%%%% 
%\begin{acknowledgments} 
%\end{acknowledgments} 
%%%%%%%%%%%%%%%%%%%%%%%%%%%%%%%%%%%%%%%%%%%%%%%%%%%%%%%%%%%%%%%%%%%%%%%%%%%%%%
%\bibliographystyle{apsrev}
%\bibliography{fp}


%%%%%%%%%%%%%%%%%%%%%%%%%%%%%%%%%%%%%%%%%%%%%%%%%%%%%%%%%%%%%%%%%%%%%%%%%%%%%%
%%%%%%%%%%%%%%%%%%%%%%%%%%%%%%%%%%%%%%%%%%%%%%%%%%%%%%%%%%%%%%%%%%%%%%%%%%%%%%
%%%%%%%%%%%%%%%%%%%%%%%%%%%%%%%%%%%%%%%%%%%%%%%%%%%%%%%%%%%%%%%%%%%%%%%%%%%%%%
\end{document}